# Space-Based Infrared Near-Earth Asteroid Survey Simulation

**Edward F. Tedesco**[*], TerraSystems, Inc., Lee, NH 03824, USA,
Karri Muinonen, Observatory, University of Helsinki, Helsinki, Finland, and
Stephan D. Price, Air Force Research Lab. (VSBC), Hanscom AFB, MA 01731, USA

*We demonstrate the efficiency and effectiveness of using a satellite-based sensor with visual and infrared focal plane arrays to search for that subclass of Near-Earth Objects (NEOs) with orbits largely interior to the Earth's orbit. A space-based visual-infrared system could detect approximately 97% of the Atens and 64% of the IEOs (the, as yet hypothetical, objects with orbits entirely Interior to Earth's Orbit) with diameters greater than 1 km in a five-year mission and obtain orbits, albedos and diameters for all of them; the respective percentages with diameters greater than 500 m are 90% and 60%. Incidental to the search for Atens and IEOs, we found that 70% of all Earth-Crossing Asteroids (ECAs) with diameters greater than 1 km, and 50% of those with diameters greater than 500 m, would also be detected. These are the results of a feasibility study; optimizing the concept presented would result in greater levels of completion. The cost of such a space-based system is estimated to be within a factor of two of the cost of a ground-based system capable of about $21^{st}$ magnitude, which would provide only orbits and absolute magnitudes and require decades to reach these completeness levels. In addition to obtaining albedos and diameters for the asteroids discovered in the space-based survey, a space-based visual-infrared system would obtain the same information on virtually all NEOs of interest. A combined space-based and ground-based survey would be highly synergistic in that each can concentrate on what it does best and each complements the strengths of the other. The ground-based system would discover the majority of Amors and Apollos and provide long-term follow-up on all the NEOs discovered in both surveys. The space-based system would discover the majority of Atens and IEOs and provide albedos and diameters on all the NEOs discovered in both surveys and most previously discovered NEOs as well. Thus, an integrated ground- and space-based system could accomplish the Spaceguard goal in less time than the ground-based system alone. In addition, the result would be a catalog containing well-determined orbits, diameters, and albedos for the majority of ECAs with diameters greater than 500 m.*

---

[*] Corresponding author. Phone +1-603-659-5620, Fax +1-435-203-8391, e-mail etedesco@terrasys.com

# Space-Based Infrared Near-Earth Asteroid Survey Simulation

Note: Click on a section in the Table of Contents to go directly to that section. The program's back button can then be used to return to this page. The same is true for links within the body of the document. When using Adobe Acrobat, links to external web pages will open in Acrobat or the default browser (your choice).

# Table of Contents

E.F. Tedesco, K. Muinonen, and S.D. Price

**Introduction**

For at least the past ten years discussions regarding searches for Near-Earth Objects (NEOs) have concentrated on the "hazard" aspect of these objects (*cf., Hazards due to Comets and Asteroids*, 1994, T. Gehrels, Ed., The Univ. of Arizona Press, a 1,300 page book devoted solely to this aspect of asteroid studies.). While the present work is certainly relevant to the hazard issue, it is intended to address the broader topic of NEO studies, including the determination of diameters and albedos for a significant fraction of the population.

We present the results of a feasibility study on the efficacy of a space-based Infrared search for, and characterization of, NEOs analogous to the ground-based visual search study by Bowell and Muinonen (1994). In particular, we employ the identical orbital and size-frequency distribution of the Earth-Crossing Asteroid (ECA) population used in that study, with the addition of a population with orbits smaller than the Earth's. The term, "Interior to Earth's Orbit" (IEO), is one coined by Michel *et al*. (2000) to describe this, as yet hypothetical, asteroid population with orbits entirely interior to that of the Earth's orbit.

The asteroid component of the NEOs has traditionally been divided into the Aten, Apollo, and Amor classes, based upon their current osculating orbital elements (Shoemaker *et al*., 1979). And, as noted above, the IEO class was added this year. IEO asteroids have semimajor axes (a) < 1 AU and aphelion distances (Q) < 0.983 AU and so are currently completely interior to Earth's orbit. Aten asteroids have a < 1 AU and Q > 0.983 AU, Apollo asteroids have a ≥ 1 AU and perihelion distances (q) ≤ 1.017 AU, while Amor asteroids have a > 1 AU, 1.017 < q ≤ 1.3 AU, and are therefore currently completely exterior to Earth's orbit.

Comets, in particular long-period comets, are not dealt with in this paper. Note, however, that even slightly active comets are readily detected at mid-infrared wavelengths as demonstrated by IRAS' discovery, during its ten-month mission in 1983, of 2 periodic and 6 non-periodic comets (B. Marsden, personal communication) and numerous comet debris trails (Sykes and Walker, 1992).

In this paper, we use the term Near-Earth Objects (NEOs) to refer to those asteroids and comets (both active and dormant) with perihelion distances less than 1.3 AU. Thus, an IEO, once found, will automatically be an NEO.

As earlier (Bowell and Muinonen, 1994 and Muinonen, 1998), we adopt the working definition that all Apollo and Aten asteroids, as well as Amor asteroids with perihelion distances less than 1.13 AU, are ECAs, and can currently, or some time in the future, intersect the capture cross section of the Earth. At this time, we treat the IEOs separately from the others. The simulations we present here use the ECA subset of the NEO population.

The actual population (numbers, orbital elements, and size distributions) of Atens and IEOs is poorly known. However, Michel *et al*. (2000), suggest that the half-life of Atens may be higher than that of the general NEO population due to the fact that protection mechanisms against close approaches occur more frequently for objects in Aten orbits. Nearly 60% of the Aten orbits in the Michel *et al*. study evolved into objects with orbits entirely inside Earth's orbit, *viz*., a < 1 AU and Q < 0.983 AU. Aten-like orbits are a "transient stage", for the transfer of Apollo objects to and from IEO objects.

# Space-Based Infrared Near-Earth Asteroid Survey Simulation

Wetherill (1979) suggested that IEO objects (although he did not call them that) should comprise between 1% and 3% of the NEO population and Michel *et al's*. (2000) preliminary conclusion is that they are of order 1% to 5% of the total population.

Michel *et al*. (2000) note that because nearly 60% of the integrated Aten orbits become Apollos confirms that there exists a quasi-continuous interchange between the Aten and Apollo groups and that their definition is purely arbitrary (a point also made by Boattini and Carusi, 1998); it is based on their current orbits, which can change on a very short time scale. IEO orbits appear to be more "stable" even though they too ultimately are strongly perturbed. Thus the Aten population is probably larger than previously thought and the IEO population about 60% that of the Atens.

Recently, there have been a number of studies (Morrison, 1992; Shoemaker, 1995; Harris, JPL, 1998) that discuss the best methods for discovering and obtaining orbits for a significant fraction of the larger NEOs.

The Shoemaker Report (Shoemaker, 1995) is a response to a request to NASA from the U.S. Congress' Committee on Science, Space, and Technology for a program plan to carry out a comprehensive ten-year survey of near-Earth asteroids and comets along the lines of the earlier recommendations of the Spaceguard Survey Working Group (Morrison, 1992). The Shoemaker Report summarizes the conclusions of that study, one of which is that: "*The most important lesson which emerges from this analysis is that the best survey strategy is to cover the entire accessible sky every month, sacrificing whatever magnitude limit is necessary to accomplish this. A very positive result is that if that strategy is followed, adopting reasonable and even conservative limit on sky observability, it is possible to obtain reasonable completeness in ten years, including objects which never quite reach out to the orbit of the Earth and hence never come to opposition. Thus the ability to observe closer to the sun or to remove horizon limitations is not a sufficient justification in itself to move to a space-based survey system.*" However, the report states elsewhere that: "*Note Added in Press by Near-Earth Objects Survey Working Group on 30 June 1995: This study was conducted under severe time restrictions, thus it was not possible to consider certain aspects that would otherwise have been explored more thoroughly. In particular, the Working Group did not consider surveys from space. Proper study of space-based surveys would have required more resources and time than were available to the Working Group, and are therefore deferred to a future study.*".

The Shoemaker Report presents the results for a simulation of an "all-sky" search, *i.e*., 20,000 sq deg of sky centered on the opposition point and searched monthly to a limiting visual magnitude of 20; See Harris (JPL, 1998) for further details. The Morrison Report presented results for a simulation of a deeper search area (a 6,000 sq deg area they referred to as the Standard Survey Region), searched monthly to a limiting visual magnitude of 22.

In this paper we will discuss the advantages offered by a space-based infrared-visual system in the search for that subset of ECAs with semimajor axes comparable to that of the Earth's (*i.e*., IEOs and Atens) and for obtaining diameters and albedos for these, and Apollo and Amor asteroids as well.

E.F. Tedesco, K. Muinonen, and S.D. Price

**The Space-Based IR Survey Simulator**

The current simulation software has been further developed from that used in Bowell and Muinonen (1994) and Muinonen (1998). There are two major new features: the software has been extended from simulations of visual surveys to thermal infrared surveys and from accounting for Earth-crossing asteroids to account for IEOs as well.

The software allows the simultaneous simulation of several different surveys (*e.g.*, visual, infrared) in space. It uses the Jet Propulsion Laboratory ephemerides (DE405; Standish *et al.*, 1997) to determine the position and motion of the Earth with respect to the Sun at a given time. The software also utilizes parts of the asteroid orbit determination software OrbFit1.1 (the newest version of this free software is available from Andrea Milani, University of Pisa, Italy at ftp://newton.dm.unipi.it/pub/orbfit).

For each search region, specified as ranges in opposition-centered ecliptic longitude ($\lambda_{OC}$), or, equivalently, in Sun-centered ecliptic longitude ($\lambda_{SC}$), and ecliptic latitude ($\beta$), the infrared space-based survey simulation software accounts for the limiting magnitude and sky-plane coverage. For the purposes of the simulation, we have positioned the space-based system at the center of the Earth.

The survey simulation begins by generating the orbital elements, diameter, and geometric albedo for a sample asteroid, and then computes the orbital evolution in the two-body approximation for the specified survey dates and search regions, registering whether the asteroid was observable or not. In order to guarantee sufficient statistics within reasonable CPU time frames, we have scaled the absolute asteroid numbers for the simulations, always using 10,000 sample asteroids in each simulation run.

The debiased ECA orbital element distribution derives from Rabinowitz *et al.* (1994). As in Bowell and Muinonen (1994) and Muinonen (1998), we have imposed additional constraints on the orbital elements generated using the model by requiring that the semimajor axes, a, be < 4.0 AU, the perihelion distances, q, be < 1.13 AU (to exclude non-Earth-crossing Amors) and the aphelion distances, Q, be > 0.983 AU. Thus the ECA population used here is a smaller, well-defined subset of the overall NEO population.

We have utilized a simple orbital element distribution for IEOs. First, we assumed that the inclination distribution of IEO asteroids is identical to that of ECAs. Second, the IEO semimajor axes are taken to be uniformly distributed between a lower bound of the perihelion distance $q_{min}$ = 0.3 AU (a practical limit, roughly the perihelion distance of Mercury) and an upper bound of the aphelion distance $Q_{max}$ = 0.983 AU. Third, having generated the semimajor axis a, we have generated the eccentricity using a uniform distribution between 0 and Min($Q_{max}$/a - 1, 1 - $q_{min}$/a).

The size distribution of the model ECA population derives also from Rabinowitz *et al.* (1994) and, except where noted, contains equal numbers of S- (assumed geometric albedo 0.155) and C-class (assumed geometric albedo 0.05) asteroids larger than a specified diameter. The reduced estimate of the number of near-Earth asteroids by Rabinowitz *et al.* (2000) affects the absolute number of such objects but not so much the size-frequency distribution or the completeness values in this, or previous, studies. For the IEOs, we have simply assumed the same size distribution, S:C ratio, and geometric albedos.

# Space-Based Infrared Near-Earth Asteroid Survey Simulation

*Computing the Infrared Brightness*

We used the Standard Thermal Model (STM - Lebofsky, *et al*., 1986) to calculate the expected flux for the model NEOs. The STM is known to provide accurate predictions for asteroids with diameters larger than about 10 to 20 km (Tedesco, 1994). For example, the 8.5 µm mag obtained by the Midcourse Space Experiment (MSX – Mill *et al.*, 1994) for the 33 km main-belt asteroid 923 Herluga was 6.0 while that predicted using the STM is 5.9 (*cf.*, Figure 5).

Nevertheless, it has been known for over twenty years that the STM overestimates the thermal flux for some small asteroids (Lebofsky, *et al.*, 1978; Lebofsky, *et al.,* 1979*)*. The small asteroids in question all happen to be NEOs but this is merely due to the fact that these are the only few-km-diameter asteroids observable to date and likely has nothing to do with any intrinsic differences between NEOs and main-belt asteroids of comparable size.

Harris (DLR, 1998) published a Near-Earth Asteroid Thermal Model (NEATM) in an attempt to improve the thermal modeling of these small asteroids. The NEATM was developed primarily to derive albedos and sizes for near-Earth asteroids using spectrophotometric data covering a broad spectral range around the ~10 µm thermal peak. The so-called "beaming" parameter, $\eta$, is varied to obtain the best fit of the model to the data. The beaming factor depends on such parameters as the thermal characteristics and rotation rate of the asteroid and the best-fit value of this parameter normally lies in the range 1 to 2 for the near-Earth asteroids in Harris' sample. Inverting the procedure to calculate the expected fluxes from a near-Earth asteroid using the NEATM therefore requires prior knowledge of the appropriate beaming parameter. Harris (DLR, 1998) suggests 1.2 as a default value but this is based on observations of only a handful of near-Earth asteroids. The STM uses $\eta = 0.756$, which was derived from observations at 10 µm of the main-belt asteroids 1 Ceres and 2 Pallas.

Compared to the NEATM, the STM gives larger fluxes at wavelengths shortward of the thermal peak. In practice, however, the difference in the fluxes predicted by the two models in the 6 to 10 µm range is on the order of 10% (*cf.*, Figure 7). This value is obtained by examination of Harris (DLR, 1998), Figs. 1 and 2, each of which presents model fits to three near-Earth asteroids. In Harris' Fig. 1, there are between five and ten data points at wavelengths between about 4 µm and 20 µm while in his Fig. 2 there are only two data points at wavelengths of 10 µm and 20 µm. For Fig. 1 the mean difference in the model flux densities at 8.5 µm between the NEATM and STM is about 3% while the mean difference obtained from Fig. 2 is around 12%. Because our proposed system operates at 8.5 µm, we decided to use the simpler STM for the purposes of flux calculations.

*Simulation Comparisons*

To verify that the results from our simulation are in agreement with previous studies (Morrison, 1992, Bowell and Muinonen, 1994, Shoemaker, 1995, and Harris (JPL, 1998) we ran it using similar assumptions to those they did. But, we ran it separately for the ECAs, Atens, and IEOs. (Note that the ECAs include Atens but not IEOs.)

E.F. Tedesco, K. Muinonen, and S.D. Price

Table 1 presents the results of this comparison. In each case, we assumed, as in most previous studies, that half the population has low albedos and half moderate albedos, that any asteroid with a visual magnitude brighter than the survey limiting magnitude would be detected, and that there were no losses (*e.g.*, those caused by image trailing, moonlight, or the galactic plane). In reality, many asteroids brighter than the survey limiting magnitude would not be detected due, primarily, to trailing losses caused by their motions and seeing effects. As noted in the previous studies, this means that the completeness limits predicted by these visual wavelength simulations are upper limits.

**Table 1. A Comparison of Visual Ground-Based Survey Simulation Results[1].**

| Survey Region | $V_{lim}$ | Previous Study | Present Study | Remarks |
|---|---|---|---|---|
| | | (% Completeness) | | |
| Whole-Sky (4π str) | 20.0 | 92 | 89 | Bowell and Muinonen (1994). |
| All-Sky (2π str) | 20.0 | 85 | 80 | Harris (JPL, 1998). |
| 12,000 sq deg (12-20) | 20.4 | 76 | 71 | Harris (JPL, 1998). Fig 3: "65° Elong" |
| Standard Survey Region (SSR) | 22.0 | 91 | 93 | Morrison (1992). After 25 yrs. A monthly survey of 6,000 sq deg to V = 22 A single average albedo was assumed and trailing losses were accounted for. |

[1]For NEOs with diameters greater than 1 km and minimum orbital intersection distances < 0.05 AU or, for Morrison (1992), Bowell and Muinonen (1994) and the present study, ECAs, and, except for the Standard Survey Region, all after ten years of surveying the specified region once per month.

Given that the "noise" in a given percent completeness result is about ±1 and that slightly different orbital populations, assumed albedos and visual phase function slope parameters were used in the various studies, the agreement is satisfactory.

The results from Table 1, but for a period of 20 years, are presented in Figure 1.

Results from other simulations we ran are in agreement with the conclusions of the Shoemaker Report. That is, we find that in ten years a monthly survey of 20,000 sq deg to a limiting V magnitude of 20, the so-called "20-20 survey", would find 77% of ECAs with diameters greater than 1 km and 53% of those with diameters greater than 500 m.

Note that the Harris (JPL, 1998) and Shoemaker Report studies considered only those asteroids whose present orbits come within 0.05 AU of the Earth's orbit because objects in such orbits could, during the next century or so, evolve into Earth-intersecting orbits. However, Harris notes that this constraint: "… does not affect the results much, compared to using a distribution of orbits not so constrained, and otherwise matching the same NEA orbit statistics.". It is these not so constrained orbits that are used in our simulation.

# Space-Based Infrared Near-Earth Asteroid Survey Simulation

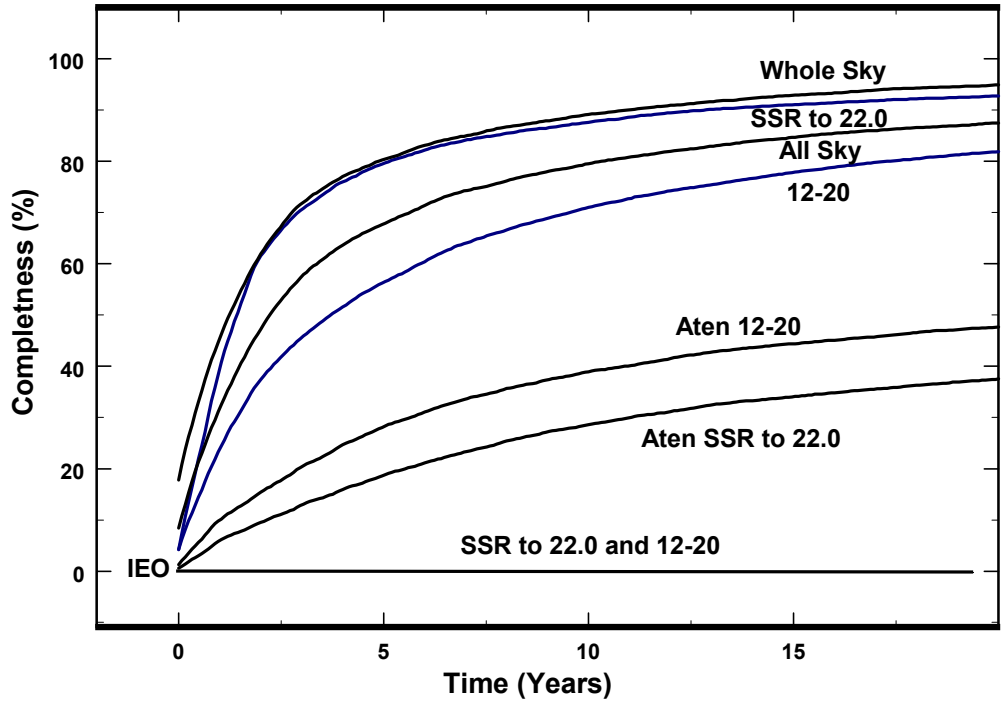

**Figure 1. Visual-Survey completeness limits as a function of time for a limiting diameter of 1 km. The top four curves are described in Table 1. The two middle curves are the completeness curves for Atens from 12,000 sq deg/month to V = 20.4 (12-20) and 6,000 sq deg/month to V = 22 (SSR) surveys. No IEOs are found in the latter two surveys.**

    Our results also confirm the statement in Shoemaker, 1995 that: "… *it is possible to obtain reasonable completeness in ten years, including objects which never quite reach out to the orbit of the Earth and hence never come to opposition*." Since according to our simulation, in ten years the idealized (*i.e.*, all asteroids with V ≤ 20 are detected and there are no other losses) 20,000 sq deg/month to V = 20 (20-20) survey would find 66% of Atens with diameters greater than 1 km and 61% of those with D > 500 m. Note that, especially where Atens are concerned (*cf.*, Figure 1), our results are in agreement with the Shoemaker Report conclusion that it is better to search more sky to a brighter limiting magnitude than a smaller area to a fainter limiting magnitude.

    Hence, the Shoemaker report statement that: "*Thus the ability to observe closer to the sun or to remove horizon limitations is not a sufficient justification in itself to move to a space-based survey system.*" is only correct if one considers finding, at best, about two-thirds of the larger members of the Aten population to be "reasonable completeness". Note also that simply searching longer does not easily raise these

completeness levels. For example, if the 20-20 survey were to continue for 100 years the Aten completeness limit for 1 km objects would not exceed 80% (or 75% for 500 m).

Furthermore, according to G. Williams (personal communication), as of early 2000 the best the worldwide search programs have done is approximately 16,000 sq deg/month, to a point-source limiting magnitude of V = 19. Williams notes that this was an exceptional month, but will almost certainly become the norm now that LINEAR started operating their second telescope (Elowitz *et al*., 2000).

With two LINEAR telescopes operating, the exposure times have been increased so that the same sky coverage will be maintained but the point-source limiting magnitude should increase to about V = 20. Only very small areas of sky are being searched to fainter than V = 20, a situation that, it appears, will continue for the foreseeable future.

During the five-year period between 1995 and 1999 a total of 38 (of the 58 known) Atens were discovered. Six of these had D > 1 km and 15 D > 500 m (assuming all had albedos of 0.155). Four of those with D > 1 km, and 5 with D > 500 m, were discovered during 1999, a year that saw the 20-20 survey goal nearly met. However, the 20-20 visual simulation predicts that 20 Atens with D > 1 km and 64 with D > 500 m should be discovered in the first year. Even allowing for the recent downsizing in the population by about a factor of two (Rabinowitz *et al*., 2000), the actual discovery rate is only about 40% and 16% of that expected for 1 km and 500 m Atens, respectively. This suggests that, at the current discovery rate, the actual time required to find two-thirds of the Atens with diameters greater than 1 km is about 25 years while about 60 years will be required to find 40% of those with diameters greater than 500 m. This estimate is consistent with Rabinowitz *et al's.* (2000) conclusion that, as of early 2000, it will take about $15 \pm 10$ years to meet the Spaceguard goal for 1 km ECAs.

Our idealized 20-20 ground-based visual search simulation predicts that about 10% of the 1 km and 8% of the 500 m IEOs would be found in ten years.

We also ran our simulation for a 16,000 sq deg/month to a point-source limiting V magnitude 20.0 to 20.5 (16-20) survey (the eventual sensitivity limit of the LINEAR systems – Viggh *et al*., 1999) on the assumption that this might be the norm for the foreseeable future. For reasons discussed in the next section (*cf*., Figure 2), for this simulation we assumed that all ECAs brighter than V = 20.0 would be detected and none fainter than this would be discovered. In this case, about 60% of the ECAs with diameters greater than 1 km would be discovered in five years, about 75% in 10 years, and 85% in 20 years. For a 500 m limit the 5/10/20 year completeness limits are about 35%, 50%, and 65%. Because the search area in our simulation covered ±110° in opposition-centered ecliptic longitude, *i.e*., to within about 70° solar elongation at the ecliptic plane, and ecliptic latitudes from –15° to +85° (to account for the fact that all the ground-based search telescopes are located at mid-northern latitudes, although this turned out to have little effect on the final results) about ten percent of the IEOs with diameters greater than 1 km might be discovered in the first ten years and 13% after 20 years. For a 500 m diameter limit, these values are 7% and 10%, respectively. These results should also be considered to be upper limits for the reasons given above.

Hence, if one wishes to obtain more than 85% completeness for D ≥ 1 km ECAs (80% for Atens and 13% for IEOs) in less than 20 years, then the ability to observe closer to the Sun is a sufficient justification in itself to move to a space-based survey system. However, as we will show, it is not the only, or even the most important, reason for going into space. Before doing so, however, we will briefly discuss the prospects for ground-based surveys specifically for Atens and IEOs.

**Ground-Based Aten and IEO Searches**

Drummond *et al.* (1993) and Boattini and Carusi (1998) discuss the problem of discovering Atens and objects with orbits interior to the orbit of the Earth. Only ground-based searches were considered. Drummond *et al.* suggested that if searches were made closer to the Sun enough objects could be discovered so that, with some reasonable assumptions, the biases could be deduced and that this would then lead to a better understanding of the dynamics that produce the observed NEO distribution. Boattini and Carusi concluded that searches should concentrate on the two regions either side of the Sun between solar elongations of 50° and 120°.

Most recently Bottke, *et al.* (1999) and Morbidelli, *et al.* (1999) addressed the issue of biases in the orbital and size-frequency distributions of NEOs. They considered only ground-based searches and, as is generally done, used absolute magnitudes, rather than diameters, to parameterize the sizes. IEOs were not treated.

We believe that the discussion by Whitely and Tholen (1998) on extending a pilot ground-based search for Aten asteroids from a 2.24-m telescope to a 3.6-m telescope with a larger and more sensitive CCD camera demonstrates the difficulty of discovering a significant fraction of even this more easily detected population, with respect to IEOs. This is primarily due to the small amount of time available each night (~4 hrs) during which regions of sky near 75° of the Sun can be searched.

Whitely and Tholen (1998, page 165) noted that, using the University of Hawaii's 2.24-m telescope with an 8k square CCD having a 19 arcminute field of view: "*For a search region near 75° solar elongation, we have managed to double scan about 1.5 square degrees of sky to a depth of V = 22.5 in about 2 hr.*". Scaling this to the 3.6-m Canada-France-Hawaii Telescope (CFHT) mosaic array (CFH12K - 12288 x 8192 pixel, 42 x 28 arcmin FOV at the CFHT prime focus), they estimate, will enable about five square degrees of sky to a depth of V = 24 to be covered in about two hours. Thus, twelve nights a month of cloudless, good seeing, dark time on the CFHT would allow about 120 square degrees to be searched to a point source limiting magnitude as faint as V = 24.

Repeating our simulation for 120 sq deg/month for opposition-centered ecliptic longitudes of 100° to 110° (corresponding to solar elongations of approximately 70° to 80°) and within 3° of the ecliptic plane to a limiting visual magnitude of 24 shows that, in a ten-year period, an upper limit of about 25% of the Atens, and 4% of the IEOs, with diameters greater than 1 km, would be discovered.

Doubling the search area to 240 sq deg/month (by increasing the range of ecliptic latitudes to ±6°) and maintaining the limiting visual magnitude of 24 results in a ten-year 1 km completeness upper limit of about 41% for the Atens, and 6% for the IEOs. Even in this idealized case, it would take 100 years to reach 1 km diameter completeness levels of 68% for the Atens and 12% for the IEOs.

As in the case of the ground-based NEO surveys, higher discovery efficiency is achieved by searching more sky to a lower limiting magnitude. For example, if 1,200 sq deg/month (opposition-centered ecliptic longitudes of 95° to 115°, *i.e.*, solar elongations of approximately 65° to 85°, and ecliptic latitudes within 15° of the plane) were searched to a true limiting visual magnitude of 20 then, in ten years, about 63% of the Atens, and 17% of the IEOs, with diameters greater than 1 km, would be discovered. As in the case of the "all sky" surveys, the actual discovery rate would be considerably less because of losses due to image trailing, weather, and Galactic plane avoidance.

Due to the presumably steep size-frequency distributions of the Aten and IEO populations, it is likely that searches such as those being conducted by Whitely and Tholen will discover small (in the ten to few hundred meter diameter size range) members of both classes. However, it is our conclusion that the discovery of significant fractions of these populations, sufficient to obtain reliable estimates of their orbital and size-frequency distributions, cannot be done on reasonable (*i.e.*, a few decades) time scales.

*Limiting Magnitudes*

In our simulations, for both the visual wavelengths presented above and the infrared wavelengths discussed below, we define "limiting magnitude" to be that magnitude below which all asteroids in the sensor's field of view will be detected. This differs from the definition used by some ground-based survey systems. For example, Viggh *et al*. (1999) state: "*We estimate the sensitivity achieved by LINEAR is approximately 19.5 visual magnitude when using 10 second integration period during good seeing. However, since the actual sensitivity achieved is a complex function of many variables, the actual performance of the search provides the best determination of the sensitivity. Figure 10 contains a histogram of 137 NEOs discovered by LINEAR, binned by their magnitude at discovery. The discovery magnitude is determined by taking the magnitude estimates made by other astronomers in the follow up process and combining them with a knowledge of the orbit to determine the apparent brightness as of the discovery date. Figure 10 indicates that the LINEAR system achieves sensitivity of ~19.5 visual magnitude as expected.*" The data for NEOs with V > 15 at discovery from Viggh *et al.'s* Figure 10 are reproduced here as Figure 2.

This difference in definitions needs to be kept in mind when comparing results from idealized simulations with those expected from actual observing programs. Nevertheless, the results presented here from the idealized simulations serve as useful upper limits on what is to be expected from actual ground-based surveys.

**Space-Based Infrared Near-Earth Asteroid Survey Simulation**

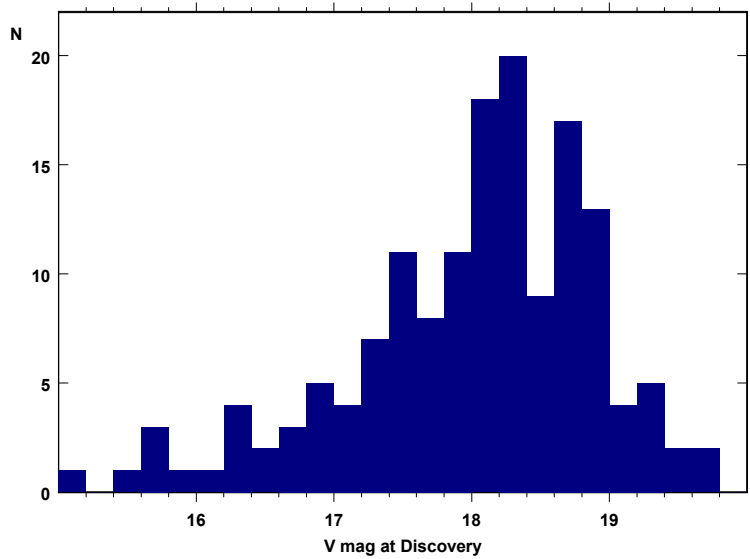

**Figure 2. Apparent brightness of LINEAR NEOs at discovery. (From Viggh *et al*., 1999, Fig 10.)**

   The turnover in Figure 2 near V = 18.3 is not real and indicates that most of the NEOs with apparent visual magnitudes greater than 18 are not being detected. The visual magnitude at discovery histogram for the 20-20 simulation discussed above is given in Figure 3. Note that there are 3.2 times as many NEOs with V at discovery in the 18.3 magnitude bin than in the 19.5 mag bin compared with a ratio of 0.1 for the LINEAR data. Thus, it would appear that LINEAR may be detecting as few as 1 NEO in 32 with V near 19.5.

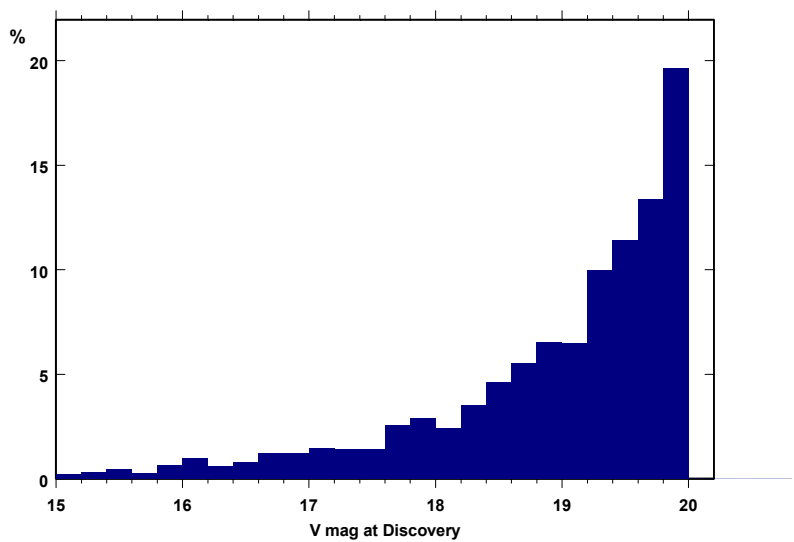

**Figure 3. Apparent brightness of 20-20 simulation NEOs at discovery.**

E.F. Tedesco, K. Muinonen, and S.D. Price

**The Need for Physical Observations**

Through the end of 1999 there were 826 known Atens, Apollos, and Amors with q < 1.13 according to the Minor Planet Center's web site (http://cfa-www.harvard.edu/iau/lists/Unusual.html) maintained by G. Williams and B. Marsden. Of these, measured diameters/albedos are available for 63, according to the EARN web site maintained by G. Hahn (http://129.247.214.46/nea/).

Figure 4 shows the incremental and cumulative discovery, together with the relative rate at which diameters and albedos are being determined, for ECAs over the 28-year period between January 1972 and December 1999, inclusive. The dashed curves (referred to the left axis) are the cumulative discoveries and the solid curve (referred to the right axis) the percentage of the total known ECAs with a measured diameter and albedo. Note that the fraction of known ECAs with a measured diameter and albedo peaked in the mid-1980s (at about 35%) and has been declining steadily ever since (to about 7% at the end of 1999).

The discovery rate for ECAs (currently over 150 per year over all size ranges) is running away from the rate at which physical characterizations are being made.

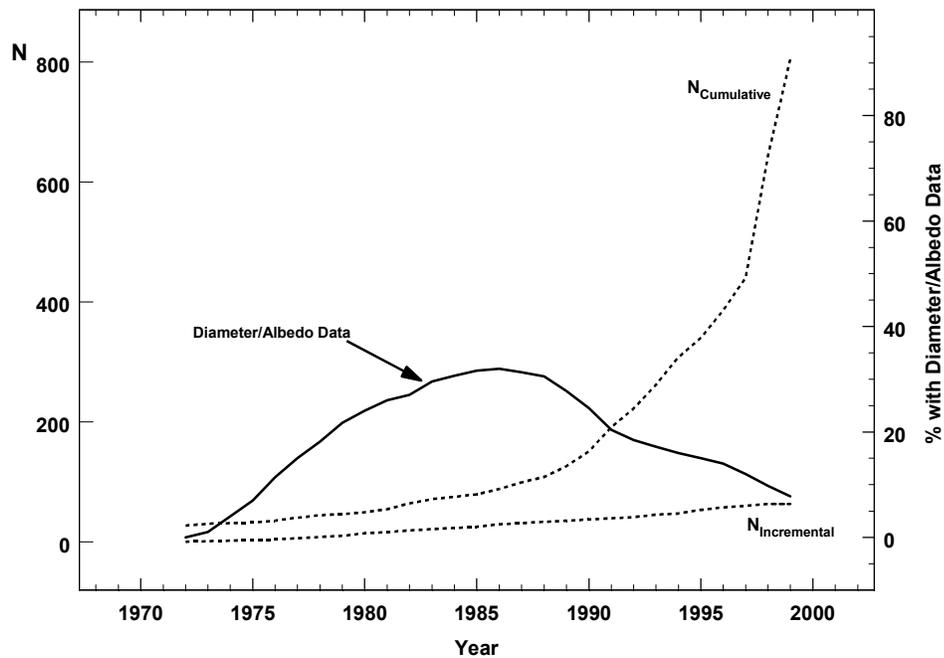

Figure 4. ECA discovery and physical characterization rates for 1972 through 1999.

# Space-Based Infrared Near-Earth Asteroid Survey Simulation

**Advantages of a Space-Based Infrared System**

Space-based instruments, in general, are subject to fewer restrictions than ground based systems. Weather and seasonal effects are eliminated and lunar constraints are much more generous, being set by the side-lobe response of the system.

The infrared is a more efficient wavelength at which to search. It is straightforward to show that the signal-to-noise ratio of a detection of an S-class ECA is about a factor of five higher in the infrared than the visible if the pixels in the CCD and mid-infrared arrays are sized to the diffraction limit of the telescope and the sensitivity of both arrays is limited by the zodiacal background. Also, unlike the linear dependence with albedo at visual wavelengths, there is virtually no dependence of the infrared flux on the geometric albedo. For example, on 21 September 2000 the visual magnitude of the Apollo asteroid 4179 Toutatis (assuming it has a diameter of 3 km and a visual geometric albedo of 0.15) is 14.4. If its albedo were 0.05 its visual magnitude would be 15.6, a difference of 1.2 mag. However, the 8.5 µm magnitudes under these same conditions, would be 3.58 and 3.53, respectively, or a difference of only 0.05 mag. Furthermore, note that the lower albedo actually results in a slightly greater 8.5 µm brightness because in the lower albedo case the asteroid would be slightly warmer.

The different ways in which the albedo affects the visual and infrared flux is significant in that, when computing survey completeness limits, assumptions regarding the albedo distribution of the target population must be made. It is usually assumed that half the population has albedos around 0.15 and half around 0.05 (a so-called "S:C Ratio" of 1). As shown in Table 2, the completeness levels reached are, in the case of visual surveys, quite sensitive to the value assumed for the S:C ratio, while in the case of the infrared, the value chosen is unimportant.

**Table 2. Percent Completeness as a Function of the S:C ratio (500 m ECAs).**

| S:C Ratio | Ground-Based 20-20 Simulation after 10 years | Space-Based IR $|\lambda_{SC}|$ 40°- 60° Simulation after 5 years |
|---|---|---|
| 0.1 | 41 | 52 |
| 1.0 | 53 | 51 |
| 10.0 | 64 | 50 |

*Unique Near-Sun Surveying Capability*

As demonstrated by the MSX detection of 923 Herluga, discovery of faint asteroids as close as 28° from the Sun is possible. MSX observed Herluga on three of the ecliptic pole-to-pole scans that used the Earth as an occulting disk to survey the Sun-centered longitudes between 25° and 30°.

923 Herluga is a 33 km main-belt asteroid with an IRAS (1992) determined albedo of 0.042. Herluga was serendipitously observed by MSX on 6, 9, and 10 August 1996 at a solar elongation of 28°, a solar phase angle of 12°, and a Galactic latitude of 10°. The heliocentric distance was 2.28 AU, the geocentric distance 3.13 AU, and the predicted V magnitude 16.5. The measured 8.5 µm magnitude was 6.0 and that predicted using the STM was 5.9. The geometry of these observations is shown in Figure 5.

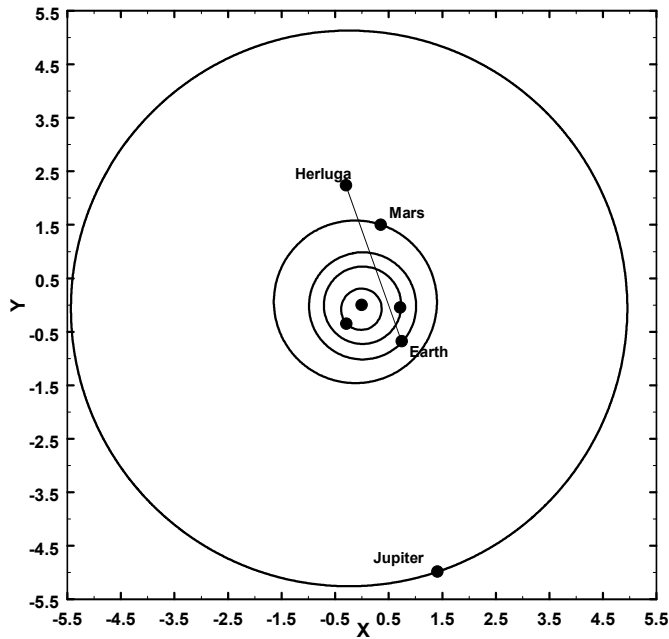

**Figure 5. Midcourse Space Experiment observation of the main-belt asteroid 923 Herluga. See text for details.**

## *Less Confusion*

There is less confusion between asteroids and the background stars at infrared wavelengths. The background sources one must contend with are much fainter in the infrared, by about 10 magnitudes (Price *et al*., 1997), than in the visible. This is because the peak flux for most stars occurs at wavelengths below 1 µm whereas that for most asteroids is around 10 µm, or longer.

Figure 6a shows the visual (E-plate) Digital Sky Survey image of the field in which MSX detected 923 Herluga. The limiting V magnitude on the original plate is about 20 (*cf*., Cutri *et al*., 1992) but probably at least 0.5 mag brighter in this reproduction. Figure 6b shows the composite MSX image (8.5 µm limiting magnitude ~ 7) of the detections on 9 and 10 Aug 1996. One infrared image is color coded green, the other red. Thus, Herluga appears as the green and red dots while stars appear yellow.

The location of the region shown is near galactic latitude 12° and galactic longitude 206°, well within the 20° latitude kept out zone adopted by ground-based surveys to avoid the Galactic plane. An 8.5 µm image with a limiting magnitude of 9 would contain approximately 120 stars and about 3 main-belt asteroids (in contrast to the visible image which contains about 25,000 stars, and 30 main-belt asteroids with V brighter than 19.5). The mid-infrared survey described herein would not have "avoidance regions", although the completeness would be compromised by confusion within 45° of the Galactic Center and within 2° of the Galactic plane. In point of fact, MSX detected asteroids within one degree of the Galactic Center (Egan *et al.*, 1998).

# Space-Based Infrared Near-Earth Asteroid Survey Simulation

Because the mid-infrared is intrinsically more sensitive for detecting NEOs than the visible, shorter integration times can be used. This, and the fact that diffraction limited mid-infrared pixels are larger, 60 - 70 µm (12" - 14") vs. visual CCD pixel sizes of 10 - 15 µm (1.0" to 1.5") means that trailing losses are negligible for the mid-infrared sensor. Thus, the sensor will detect virtually all asteroids brighter than the detection threshold. A signal-to-noise threshold of 6 is generally adopted for a single uncorrelated measurement.

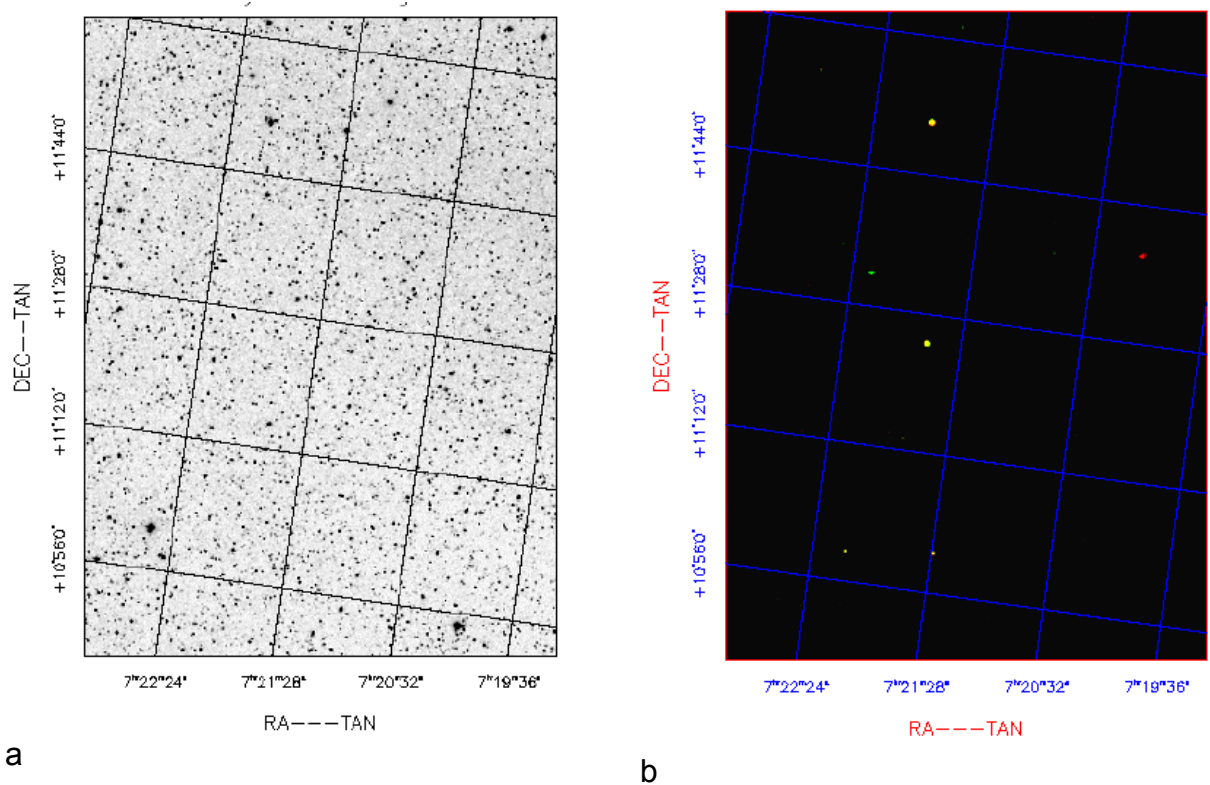

a                                                                b

**Figure 6. (a) Asteroid 923 Herluga field from a 50-minute exposure POSS E plate obtained at 3:37 UT on 23 March 1955 - a Digital Sky Survey image[1] and (b) a composite MSX image from August 1996. See text for details.**

## *Choice of Infrared Wavelength*

The optimum infrared wavelength for an NEO search is centered on the peak of the thermal emission. As can be seen from the infrared spectrophotometry of Eros in Figure 7, the mid-infrared emission peaks in the 6-to-10 µm region. The spectral energy distribution of Eros is typical of the other near-Earth asteroids studied by Harris (DLR). Thus the 6-to-11 µm band proposed for the space-based infrared survey is ideally suited to the task.

---

[1] Based on photographic data of the National Geographic Society -- Palomar Observatory Sky Survey (NGS-POSS) obtained using the Oschin Telescope on Palomar Mountain. The NGS-POSS was funded by a grant from the National Geographic Society to the California Institute of Technology. The plates were processed into the present compressed digital form with their permission. The Digitized Sky Survey was produced at the Space Telescope Science Institute under US Government grant NAG W-2166.

E.F. Tedesco, K. Muinonen, and S.D. Price

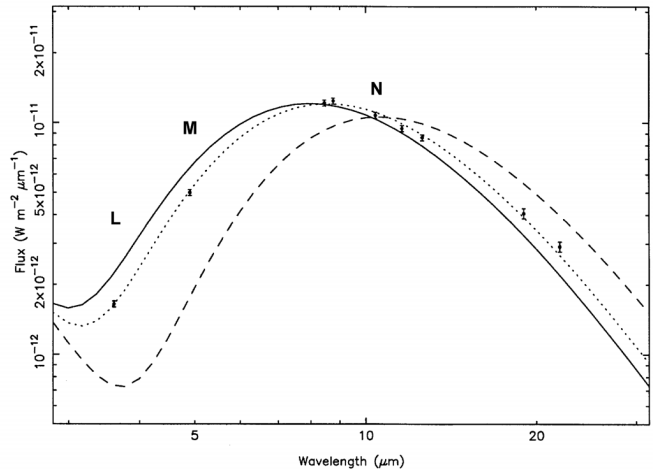

**Figure 7.** Thermal model fits to observed infrared fluxes in the range 4–20 µm for 433 Eros. Adapted from Harris (DLR, 1998) Fig. 1. The solid line is the Standard Thermal Model, the dotted line the Near-Earth Asteroid Thermal Model, and the dashed line the Fast Rotating Model.

**Straw Man Spacecraft**

The feasibility of a space-based infrared NEO survey system is proven by MSX. The survey detection threshold for the 8.5 µm band was $7^{th}$ magnitude (Price and Egan, 1999), the minimum considered in this study. This was achieved using a 13 msec sample time with an effective integration time of ~0.33 msec, the time it takes for a source to transit the eight 0.306 arc minute pixels at the survey scan rate of 7.5 arc minutes per second. If MSX had a square 192x192 array with identical performance as the 8x192 actually used, the effective integration rate would have been ~8 sec and the limiting magnitude 8.7. This gain was verified in that the 20 second effective integration time produced by adding repeated scans over an area at a slower rate did, indeed, produce the anticipated 9.2 magnitude threshold.

Most of the NEO survey objectives could be met by straightforward modifications of the MSX infrared telescope. Correcting for the cryogen lessons learned on the MSX mission (Schick and Bell, 1997) would increase the lifetime to about two years. The modeled thermal performance (Ames and Burt, 1993, 1994) budgeted half the observing time against high thermal inputs such as looking at the Earth or Earth limb. The NEO survey will only observe against the low thermal backgrounds of space. That, plus configuring the spacecraft to optimize radiative cooling of the shell, adds another year. Sample times of more than 1 second can be realized by using the MSX internal scan mirror to "freeze frame" the scan motion across a large format array. Readout and multiplexer noise dominate the MSX focal plane noise at a sample time of 13 msec. Hence, increasing the readout time to more than 1 sec improves the sensitivity by more than the $\sqrt{t}$ difference. A further gain of a factor of two-to-three would be realized using a Si:Ga array, which has lower dark current at the higher MSX focal plane temperatures than the Si:As arrays that were used. The pixel size would be reduced to 12", the **Space-Based Infrared Near-Earth Asteroid Survey Simulation**"diffraction limit" for an 8.5 µm band. Thus, most of the infrared mission objectives could be met with easily achieved changes to an existing design of the modest sized (35-cm) MSX experiment.

## *Proposed NEO Spacecraft*

Expendable cryogen is inadequate for a five-year plus mission. Thus, the proposed instrument would be a 0.6-m on-axis telescope cooled by redundant mechanical cryocoolers. Davis and Tomlinson (1998) reviewed the recent status of cryocooler development. There are several coolers undergoing bench testing that are capable of achieving the 35K required of the optics and 10-15K needed by the Si:Ga focal plane. Mechanical coolers eliminate the need for a vacuum shell and are less massive than a cryotank. Development of sensitive, large format, mid-infrared Hg:Cd:Te arrays has rapidly matured (*e.g.* Mandl and Ennulat, 1997). These focal planes have the felicitous properties of having the desired spectral response (equivalent to the MSX 8.5 µm band), large format (128x128) and a relatively high operating temperature (35K).

The baseline sensor is a 0.6-m on-axis telescope, diffraction limited at 0.8 µm with a visible CCD and a 256x256 infrared focal plane. 0.6-m optics were chosen for several reasons. As Cellino *et al.* (2000) point out; the visible CCD needs the increased collecting area for increased sensitivity. There is a corresponding increase in infrared sensitivity as well as higher spatial resolution. Finally, the aperture is the same as the telescopes on IRAS and ISO so these systems can be used as benchmarks. The satellite is launched into a 900 km altitude Sun-synchronous orbit. The infrared pixels are sized to the first Airy null and are ~7.5". The focal plane covers an instantaneous field of view of 32'x32'. The visible CCD parameters are such that it covers the same instantaneous field of view but with smaller pixels. Nominally, there is about a factor of four difference between the inter-pixel spacing (pitch) of infrared arrays and visible CCDs. An internal scan mirror, or a scanning secondary, is used to "freeze frame" the spacecraft scan, much like the secondary mirror used in the Two-Micron All Sky Survey (2MASS). A given source will, therefore, be observed on six 1.5-sec exposures if the spacecraft is scanned at the orbital rate of ~3.6 arc minutes/sec. The focal plane will be tilted slightly with respect to the scan direction to provide critical sampling in both the in-scan and cross-scan directions, again as used by 2MASS. The Sun-centered ecliptic longitudes between 40° and 60° are redundantly covered by offsetting the scan by a quarter of a degree on successive scans. The area is covered by roughly 80 scans taken over a week. The survey could be repeated on a weekly or bi-weekly basis to improve the visual and infrared radiometry and the quality of the orbits.

## *Astrometry*

The positions of the MSX extractions are accurate to about $1/8^{th}$ of a pixel and those of 2MASS to better than a tenth. Thus, the positional accuracy of the proposed system should be between 0.5" and 1" using the 2MASS catalog for the absolute reference. The proposed system will detect every 2MASS source brighter than the detection threshold, *i.e.*, several million stars. The 2MASS positions are known to better than 0.2" and there should be at least several tens of 2MASS stars in a single mid-infrared image frame.

- 18 -

MSX also flew an on-board signal processor as an experiment that successfully demonstrated the possibility to autonomously extract and quantify sources in the infrared data stream. The proposed system would use such a capability not only to identify sources but also to derive the astrometry and then to identify the location of potential NEOs in the frame. Stars are rejected through positional matches with 2MASS sources and a color criterion. Pre-launch, the color differences for the 2MASS sources will be calculated. As the survey progresses, the measured differences will be substituted and tighter acceptance bounds instituted. Once potential NEOs are identified in the frame, a message packet is extracted from the visible CCD at that location. A message packet is the output from a block of pixels somewhat larger than is expected to contain the source. At a minimum, the infrared and visible message packets should be telemetered to the ground along with the infrared position and brightness of all the extracted sources.

Based on the costs for the infrared sensor on MSX, the Wire program, and the maturity of cryocoolers and mid-infrared Hg:Cd:Te arrays, a space-based system capable of attaining the objectives presented in this paper is within the constraints of the NASA MIDEX/Discovery program, *i.e*., less than 140 million USD.

## Simulated Survey with the Proposed NEO Spacecraft

As mentioned in the Introduction, we employ the identical ECA population orbital and size-frequency distribution used by Bowell and Muinonen (1994), with the addition of the IEO population. Details on this addition and how we compute the 8.5-µm magnitude for each asteroid in the sensor's field of view are given in the section on The Space-Based IR Survey Simulator, above.

### *Infrared Sensor Sensitivity*

As we will demonstrate in the following section, to meet our objectives, the area of sky that needs to be searched at small solar elongations is significantly less than that which must be covered at large elongations. Consequently, the ultimate completeness to any given size limit is more strongly dependent upon the limiting magnitude of the survey. To determine the required limiting magnitude we ran the simulation for the 1,000 m Aten, $|\lambda_{sc}|$ 40° – 50° case described in Table 3 but for limiting 8.5 µm magnitudes of 7, 8, 9, and 10. From the results presented in Figure 8 we conclude that a limiting 8.5 µm mag of 9.0 is sufficient to meet our objectives. This is about ten times more sensitive than MSX in its survey mode. As noted above, this can be easily achieved with a modest sized infrared telescope.

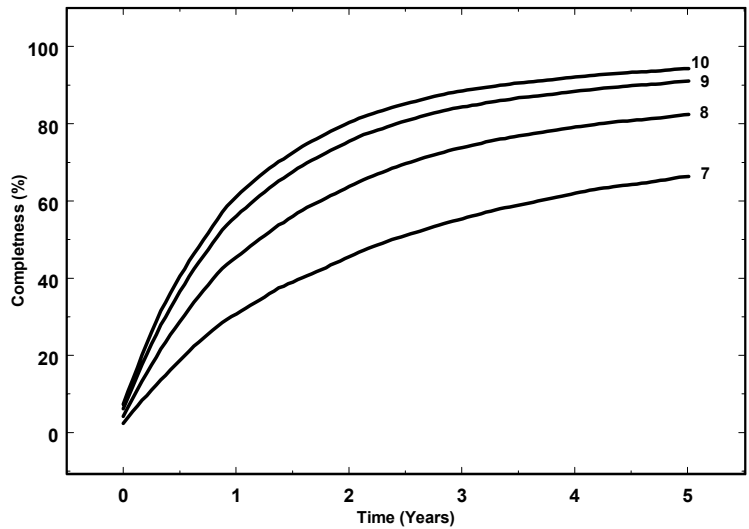

Figure 8. Percent completeness as a function of limiting 8.5 μm mag for the $|\lambda_{SC}|$ 40° – 50° case for 1,000-m Atens.

*Search Region*

As previously noted, the simulation code search regions are specified as ranges in either opposition-centered ecliptic longitude ($\lambda_{OC}$) or Sun-centered ecliptic longitude ($\lambda_{SC}$) and ecliptic latitude ($\beta$). For space-based surveys, especially at mid-infrared wavelengths, the Sun is the most significant constraint on where a space-based sensor can observe. This constraint is parameterized as the solar elongation ($\varepsilon$), *i.e.*, the angular distance between the sensor field of view and the Sun. For the infrared simulations discussed herein we describe the search regions, of which there are always two located symmetrically on either side of the Sun, in terms of $\lambda_{SC}$ and $\beta$. For $\beta = 0$, $\varepsilon = \lambda_{SC}$; for $\beta \neq 0$, generally $\varepsilon \neq \lambda_{SC}$. For the range of search regions we used in the infrared survey simulations (30° ≤ $|\lambda_{SC}|$ ≤ 60° and $|\beta|$ ≤ 25°), the difference, $|\varepsilon - \lambda_{SC}|$, (defining elongations to be negative west of the Sun and positive east) is negligible for our purposes, reaching a maximum of 6° at $\lambda_{SC}$ = 40°, $\beta$ = 25°. Thus, specifying the simulation search region in Sun-centered ecliptic longitude, rather than in terms of solar elongation, results in a slight underestimation of the number of predicted discoveries because the discovery rate for Atens and IEOs is higher at smaller solar elongations.

From purely geometric considerations it is clear that in a survey for Atens and IEOs the closer to the Sun one can observe, the better. However, thermal loading imposes constraints on the minimum possible solar elongations one can reach. Prior to MSX, infrared spacecraft maintained a keep-out zone at ~60° solar elongations. This eased the thermal design requirements and there was no compelling reason related to the mission objectives of these experiments to observe closer to the Sun. It is possible, with careful design, to lower this to 40°. And, as demonstrated by the MSX observation of 923 Herluga, it is possible to observe as close as 25° during eclipse season. We therefore ran the simulation for a range of ecliptic solar elongations ($\varepsilon_{ec}$; *i.e.*, solar elongations measured along the ecliptic) between 30° and 60° and with ecliptic latitudes up to 25° from the plane. All searches were limited to the maximum area that could be covered, to a limiting magnitude of 9.0 at 8.5 µm, using between one-third and two-thirds of the total spacecraft time. The remaining time is devoted to obtaining whatever follow-up astrometry is necessary to obtain a sufficiently accurate preliminary orbit and to obtain physical observations of previously discovered NEOs at higher solar elongations.

Table 3 presents the results for a subset of the cases we examined. All cases assume an S:C ratio of one (*i.e.*, equal numbers of objects with visual geometric albedos of 0.155 and 0.05, respectively), a limiting magnitude of 9.0 at 8.5 µm. $|\lambda_{SC}|$ = x° to y° means that two search regions, one either side of the Sun at Sun-centered ecliptic longitudes between x° ≤ $|\lambda_{SC}|$ ≤ y° are surveyed. The combined area of the two search regions follows this.

**Table 3. Completeness Limits for Representative Search Regions.**

| Class | % Completeness | | |
|---|---|---|---|
| | 1,000 m | 500 m | 200 m |
| a. $|\lambda_{SC}|$ = 30° to 50°; 1,186 sq deg twice per (lunar) month | | | |
| IEO | 77 | 68 | 42 |
| Aten | 97 | 88 | 47 |
| ECA | 64 | 44 | 17 |
| b. $|\lambda_{SC}|$ = 40° to 60°; 1,186 sq deg twice per (lunar) month | | | |
| IEO | 60 | 56 | 31 |
| Aten | 94 | 82 | 39 |
| ECA | 60 | 40 | 13 |
| c. $|\lambda_{SC}|$ = 40° to 50°; 594 sq deg twice per (lunar) month | | | |
| IEO | 59 | 50 | 25 |
| Aten | 91 | 76 | 32 |
| ECA | 51 | 32 | 10 |
| d. $|\lambda_{SC}|$ = 40° to 60°; 1,937 sq deg four times per (lunar) month | | | |
| IEO | 64 | 61 | 40 |
| Aten | 97 | 91 | 53 |
| ECA | 71 | 51 | 21 |

Increasing the time interval between searches to once per month lowers the five-year completeness by about 30%. For example, for case c the Aten 200 m completeness dropped from 32% to 23%.

Figure 9 displays results for the Aten and IEO searches summarized in Table 3, although only the results for 1,000 m and 500 m Atens and IEOs are plotted.

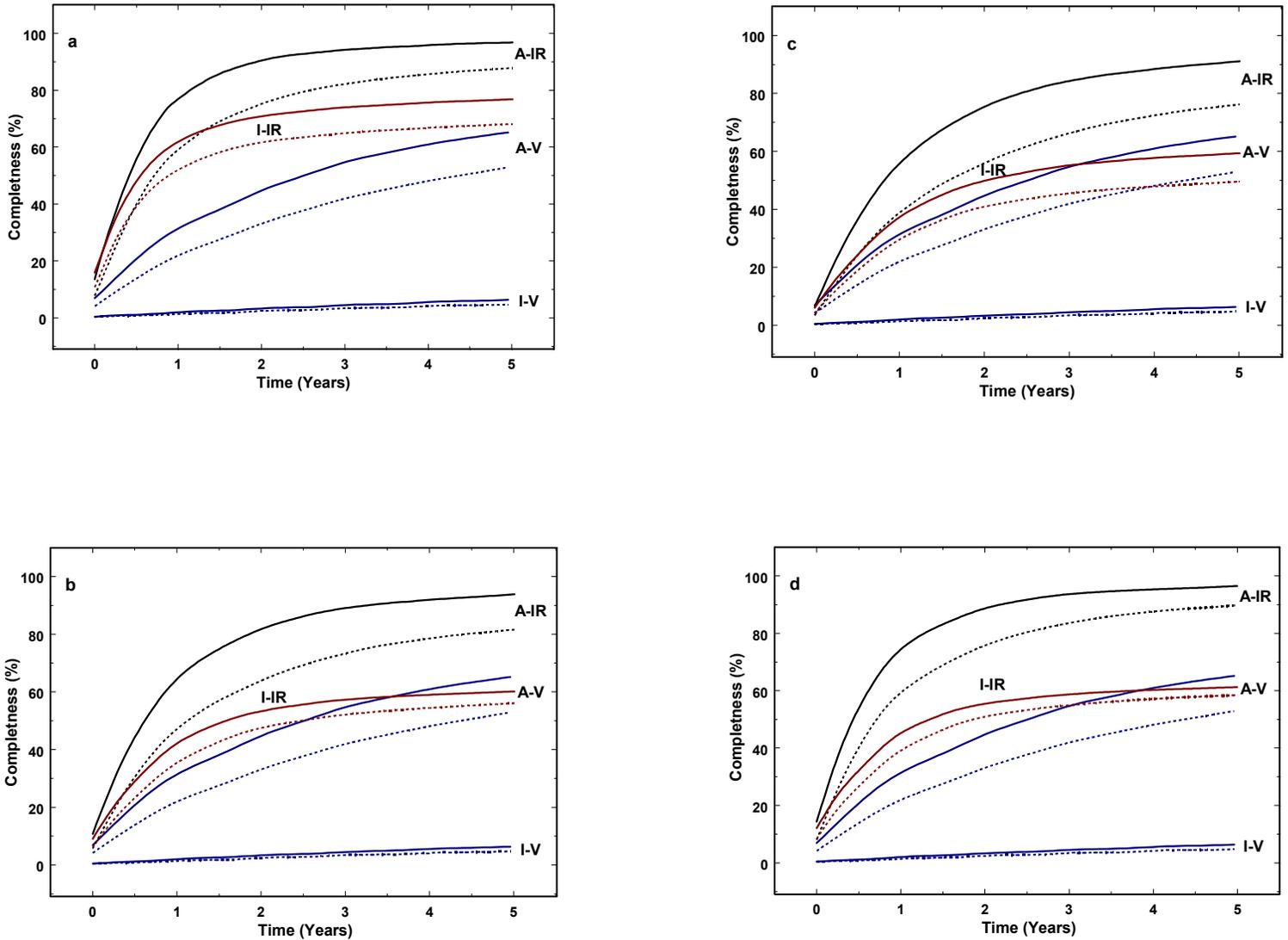

**Figure 9.** Percent of the model ECA population discovered as a function of time. Plots a, b, c, and d correspond to the search regions described in Table 2. Solid lines represent a diameter limit of 1,000 m and dotted lines a limit of 500 m. Curves labeled "A-IR" are the infrared survey simulation results for Atens and those labeled "I-IR" the results for IEOs. Results for the visual simulation (16,000 sq deg/month to V ≤ 20.0), reproduced in each of the four plots, are labeled "A-V" and "I-V". (See text for details.)

In each plot the solid lines represent a diameter limit of 1,000 m and the dotted lines a limit of 500 m. The pair of curves labeled "A-IR" in each plot gives the results for the Atens and the pair labeled "I-IR" the results for the IEOs. Results for the 16,000 sq deg/month to a limiting visual magnitude 20.0 survey, discussed in the The Space-Based IR Survey Simulator section, above, are also shown in each plot where they are labeled "A-V" and "I-V".

The visual survey results are included because they represent the best expected from ground-based surveys in the foreseeable future. In reality, for reasons previously noted, the visual curves are upper limits whereas, we believe, the infrared results are lower limits.

**Summary and Conclusions**

Table 4 compares the results for the space-based IR survey simulation (Table 3d and Figure 9d) with those for the 16-20 ground-based survey simulation. From these results we conclude that, over a period of twenty years, the ground-based surveys may discover a larger fraction of the ECA population than the space-based IR system would in five years. However, it is equally clear that in one-fourth the time the space-based IR system would find 1.2 (1.25) times as many 1,000 m (500 m) and larger Atens and about 5 (6) times as many 1,000 m (500 m) and larger IEOs than the ground-based surveys.

**Table 4. Comparison of Space-Based IR and Ground-Based Visual Survey Simulations.**

| NEO Orbital Class | IR Survey Percent Completeness after (2) and 5 Years[1] | Ground-Based Survey Percent Completeness after 20 Years[2] |
|---|---|---|
| (D ≥ 1,000 m) | | |
| ECA | (46)  70 | 86 |
| Aten | (89)  97 | 80 |
| IEO | (56)  64 | 13 |
| (D ≥ 500 m) | | |
| ECA | (31)  50 | 65 |
| Aten | (76)  91 | 73 |
| IEO | (51)  61 | 10 |

[1] $|\lambda_{sc}|$ = 40° – 60°; 1,937 sq deg; four times per (lunar) month to 8.5 µm mag $\leq$ 9.0.
[2] 16,000 sq deg/month to V ≤ 20.0

No one can say with certainty how long a ground-based system will take to reach even the limited Spaceguard goal of good-quality orbits for 90% of the NEOs with diameters greater than 1 km. However, as we have demonstrated herein, in general agreement with Rabinowitz *et al.* (2000), given current capabilities and the near-term outlook, this goal is unlikely to be reached in less than two decades. In the best case, assuming the availability of several dedicated 1-m aperture telescopes with a true

# Space-Based Infrared Near-Earth Asteroid Survey Simulation

limiting magnitude of 21.0 and operating at maximum efficiency, Harris (JPL, 1998) estimates that it would require a total of ten years to exceed 90% NEO completeness for diameters larger than 1 km. If follow-up observations are performed with the same telescopes used for the discovery survey then: "… a system to survey to 21 mag should probably consist of half a dozen 1-m telescopes." (Harris, JPL, 1998).

The same is true of ground-based searches specifically for Atens and IEOs. In theory, a ground-based search program to V = 24 could, in ten years, discover, at most, about 25% of the Atens, and 4% of the IEOs, with diameters greater than 1 km. However, that is assuming that about one-third the dark time on a 4-meter class telescope, for over a decade, is devoted to this 120 sq deg per month survey.

As noted previously, this situation can be improved for the larger Atens and IEOs by using smaller telescopes and searching 1,200 sq deg/month to V = 20. In this case, according to our simulation, upper limits of about two-thirds of the Atens, and one-sixth of the IEOs, with diameters greater than 1 km might be discovered in ten years. However, in this case, where are these telescopes to come from? Our simulation covered Sun-centered ecliptic longitudes between 65° and 85°. But the primary NEO search covers Sun-centered ecliptic longitudes primarily greater than 80°, *i.e.*, opposition-centered ecliptic longitudes less than 100°. Thus, to conduct both a general NEO survey and an Aten/IEO survey, even more 1-m class telescopes would be needed or back-to-back surveys would have to be performed.

On the other hand, conservatively, a space-based infrared system could discover 97% of the Atens, and 64% of the IEOs, with diameters greater than 1 km, and a significant fraction of 500 m Atens and IEOs (91% and 61%, respectively), in as little as five years. In addition, as a by-product of the Aten/IEO search, about 70% of the ECAs with diameters greater than 1 km, and 50% of those with diameters greater than 500 m, would be discovered, thus greatly reducing the time required to survey a significant fraction of the ECA population.

Perhaps even more importantly, a space-based system similar to that proposed herein would obtain albedos and diameters for virtually all known ECAs, both those it discovered and those discovered previously and by other means elsewhere during its lifetime. This would allow true size-frequency distributions to be derived for ECAs and provide a definitive answer to the question of the ECA albedo distribution.

We are not suggesting that the proposed visual and mid-infrared space-based survey is an alternative to ground-based surveys. Instead, we see it as a key component of an integrated hybrid ground- and space-based system that would be most efficient in terms of the numbers of ECAs discovered and characterized per unit time, and possibly in overall cost as well. The ground-based visual system would search the region within about 100° of the opposition point and follow-up all discoveries made by either system. The space-based infrared system would concentrate on searching the near-Sun sky (solar elongations less than 60°) and obtaining physical observations of virtually all the NEOs with diameters greater than a few hundred meters discovered in either the ground-based or space-based surveys. Because the sky-plane velocity of the asteroids discovered in the space-based survey would be known, follow-up ground-based observations would be more easily made, and smaller telescopes required, than if these same telescopes had to discover, follow-up, and characterize these asteroids.

Thus, for between perhaps two and three times the projected cost of the currently contemplated ground-based system an integrated ground- and space-based system could accomplish the Spaceguard goal in less time than the ground-based system alone. In addition, the result would be a catalog containing well-determined orbits, diameters, and albedos for the majority of ECAs with diameters greater than 500 m.

While the present IEOs are not a hazard the same cannot be said of the Atens. But, looking beyond the hazard issue, the large numbers of small near-Earth objects (likely including a number of relatively inactive comets) that would be discovered <u>and characterized</u> in the program described herein will provide a treasure map of easily-accessible near-Earth resources, including water, carbon, oxygen, nitrogen, iron, precious metals, and rare Earth elements. The exploitation of even one of these objects would recover the cost of this program many fold.

Finally, note that the results presented here are based upon a first-order feasibility study; the entire trade space has yet to be examined. Optimization of the system described above will undoubtedly lead to higher completeness limits for the space-based infrared survey than those presented here.

## Acknowledgments

We thank Alan Harris (DLR) for valuable discussions regarding his near-Earth asteroid thermal model, Sean Carey and Mike Egan (AFRL/VSBC) for creating the images used in Figure 6, Jukka Piironen for computer system support, and Andrea Carusi and an anonymous referee for their constructive reviews. EFT acknowledges support from the United States Air Force Research Laboratory, Battlespace Environment Division, Background Characterization Branch (VSBC), KM from the Academy of Finland, and SDP from the Ballistic Missile Defense Organization and NASA.

# Space-Based Infrared Near-Earth Asteroid Survey Simulation

E.F. Tedesco, K. Muinonen, and S.D. Price

**Figure Captions**

# List of Tables